\documentclass[10pt, a4paper]{article}
\usepackage{graphicx}
\usepackage{indentfirst,csquotes}

\usepackage{amssymb,amsthm,amsmath}
\usepackage{xcolor,paralist,hyperref,titlesec,fancyhdr,etoolbox}

\usepackage{booktabs}
\usepackage{float}

\hypersetup{ colorlinks=true, linkcolor=black, filecolor=black, urlcolor=black }

\begin{document}
\title{GluPredKit: Development and User Evaluation of a Standardization Software for Blood Glucose Prediction} 
\author{Miriam K. Wolff, Sam Royston, Anders Lyngvi Fougner, Hans Georg \\ Schaathun, Martin Steinert, and Rune Volden}
\date{\today}
\maketitle

\let\thefootnote\relax

\begin{abstract}
Blood glucose prediction is an important component of biomedical technology for managing diabetes with automated insulin delivery systems. Machine learning and deep learning algorithms hold the potential to advance this technology. However, the lack of standardized methodologies impedes direct comparisons of emerging algorithms. This study addresses this challenge by developing GluPredKit, a software platform designed to standardize the training, testing, and comparison of blood glucose prediction algorithms. GluPredKit features a modular, open-source architecture, complemented by a command-line interface, comprehensive documentation, and a video tutorial to enhance usability. To ensure the platform's effectiveness and user-friendliness, we conducted preliminary testing and a user study. In this study, four participants interacted with GluPredKit and provided feedback through the System Usability Scale (SUS) and open-ended questions. The findings indicate that GluPredKit effectively addresses the standardization challenge and offers high usability, facilitating direct comparisons between different algorithms. Additionally, it serves an educational purpose by making advanced methodologies more accessible. Future directions include continuously enhancing the software based on user feedback. We also invite community contributions to further expand GluPredKit with state-of-the-art components and foster a collaborative effort in standardizing blood glucose prediction research, leading to more comparable studies.

\end{abstract} 

\bigskip


\section{Introduction}

\begin{figure}[!ht]
    \centering
    \includegraphics[width=1\linewidth]{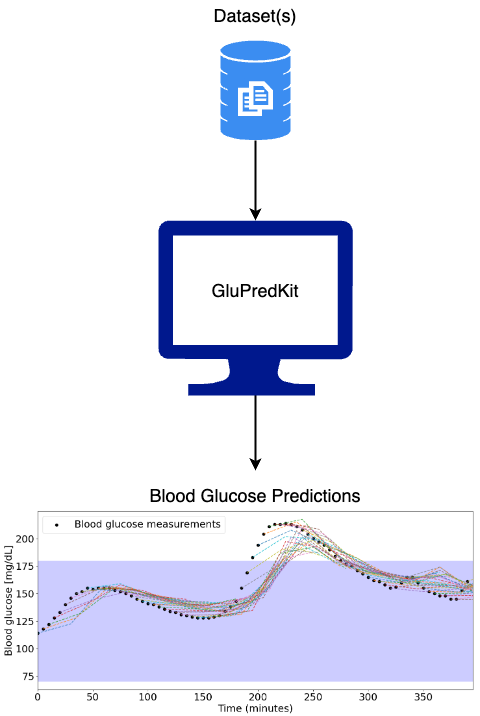}
    \caption{High-level visualization of the GluPredKit ecosystem. The upper image symbolize data storage and how GluPredKit acquires its data, while the bottom graph exemplifies the potential output of BG predicted trajectories in GluPredKit. The trajectories represent predictions for 120 minutes. Predictions are cut off when there are no more measurements to compare with.}
    \label{fig:glupredkit}
\end{figure}

\subsection{Background}
Diabetes Mellitus is a global health challenge. In 2021, there was approximately 10.5\%, or 537 million, of adults between 20-79 years living with diabetes. The International Diabetes Federation predicts that the number will increase to 12.2\% by 2045 \cite{IDF2021}. About 8 to 9 million have type 1 diabetes (T1D) \cite{gregory_global_2022}. Inadequate glycaemic control over time can damage the heart, blood vessels, eyes, kidneys, and nerves, increasing the probability of several health complications. Strengthening the glycaemic regulation for these patients could improve their quality of life, cut healthcare costs, and boost diabetics' workforce participation, highlighting the need for ongoing innovation in this field.

Recent advancements in technologies like Continuous Glucose Monitoring (CGM) devices and insulin pumps have improved glucose control, yet they require manual inputs like meal announcements and carbohydrate counting \cite{zhou_closed-loop_2022}. The challenge lies in the slow onset of subcutaneous insulin and its high variability among individuals \cite{gingras_challenges_2018}. Insulin and blood glucose (BG) dynamics are influenced by factors like meals, sleep, exercise, stress, and the menstrual cycle \cite{cinar_multivariable_predictive_control}\cite{herranz_glycemic_2016}. Non-diabetics may also exhibit significant variability in BG levels and insulin responses, implying that stabilizing glucose fluctuations could benefit them as well \cite{hall_glucotypes_2018}.

Despite technological advancements, managing diabetes remains complex. Current systems, while advanced, fall short of providing fully automated and personalized diabetes management solutions. More sophisticated BG prediction algorithms that can adapt to individual variations and real-life conditions could be used to develop enhanced artificial pancreas or decision support systems, which could alleviate the burden of diabetes management and improve patient outcomes \cite{dovc_evolution_2020}. 

There is a need for standardization in the development and evaluation of these predictive models \cite{jacobs_artificial_2023}. The current landscape is marked by diverse approaches and methodologies, leading to challenges in comparing and validating different models. Hence, the real-life applicability of models is a pressing issue. Although numerous studies claim robust algorithms for BG prediction \cite{BevanCoenen2020}\cite{zhu_blood_2020}\cite{idriss_predicting_2019}, there is a recurring limitation of validation on identical datasets \cite{hameed_comparing_2020-2}. Real-world validation of algorithms remains a significant gap in this research area \cite{diouri_hypoglycaemia_2021}. Increased use of shared code or more detailed implementation descriptions could help accelerate research in the field.

\subsection{Related Work}
\label{related_work}
Historically, BG prediction has relied on physiological models.  These models take into account various factors, including insulin absorption \cite{wilinska_insulin_2005} and meal models \cite{morch-thoresen_review_2021}. Additionally, some models for accounting for physical activity are proposed, using inputs such as heart rate or oxygen consumption \cite{jaloli_modeling_2023}. The advent of wearable sensors and smart devices has recently shifted the focus towards data-driven approaches, including machine learning and neural networks \cite{rodriguez-rodriguez_applications_2023}. 

Despite these advancements, the field faces critical model comparability and validation challenges. Jacobs et al. highlight the diversity in methodologies and datasets, which complicates direct comparisons between different studies \cite{jacobs_artificial_2023}. Initiatives such as the BG level prediction challenges in International Workshops on Knowledge Discovery in Healthcare Data (KDH) in 2018 and 2020 have addressed these challenges by providing standardized datasets and evaluation criteria \cite{MarlingBunescu2020}\cite{BGLPChallenge2020}. 

Standardization efforts, while valuable, often overlook the viability in real-life scenarios. For instance, the commonly used 30- and 60-minute prediction horizons may not capture the longer-term effects of meals and insulin. Moreover, more open code and detailed implementation information are needed to ensure progress in model validation and adaptability. A participant in the KDH workshop pointed out the restrictive nature of excluding what-if scenarios in model training and evaluations \cite{joedicke_analysis_nodate}, which are vital for developing decision-support algorithms \cite{contador_blood_2022}. In addition, what-if events are essential for model predictive control (MPC) because MPC relies on evaluating counterfactual actions in the future \cite{SpringerMPC2021}. 

Growing concern about irreproducible results across various scientific fields highlights the need for improved transparency and robustness in research methodologies \cite{Perakakis2018Open}. An open-source software allow for easy integration and comparison of different BG prediction models could encourage collaborative efforts and reproducibility and foster innovation, thereby advancing practical applications in diabetes management.

\subsection{Study Objectives}
This study aims to enhance the reproducibility, training, and comparative analysis of BG prediction algorithms. We focus on two key research questions:

\begin{enumerate}

\item What are the essential functional and non-functional requirements for a software tool in BG prediction, as indicated by existing literature?
\item How effective is the implemented software in meeting these functional and non-functional requirements, as evaluated through technical assessment and user feedback?

\end{enumerate}

GluPredKit, our open-source platform, integrates various preprocessing techniques, BG prediction methodologies, evaluation metrics, and plots. Figure \ref{fig:glupredkit} offers a high-level illustration of the GluPredKit ecosystem. The platform offers a standardized approach to facilitate data processing, algorithm training, evaluation, and comparison. It supports fetching data from prevalent diabetes data sources, ensuring simplicity when applying new data to existing model implementations. We assess GluPredKit technically against defined functional and non-functional requirements and evaluate its usability with a user study employing the System Usability Scale (SUS) and open-ended questions involving a cohort of four participants.

BG prediction models that are trained using GluPredKit can be integrated into practical applications in diabetes management, like software applications for real-time predictions \cite{wolff_mobile_sdk}, applications providing decision support, or in MPC or as an additional component for other control strategies like PID or reinforcement learning \cite{hernandez_glucose_2021}.

\subsection{Paper Structure}
The "Methods" section outlines GluPredKit’s system evaluation criteria, including the design and definition of requirements. This is followed by a description of the user study design and participant profiles. In the "Results" section, we present the technical compliance of GluPredKit with established requirements and the findings from the user study. The "Discussion" section interprets these results, considering study limitations and suggesting directions for future research. The paper concludes with the "Conclusions" section, which summarizes key findings and discusses their significance in the field of BG prediction.


\begin{figure*}[!ht]
    \centering
    \includegraphics[width=0.9\linewidth]{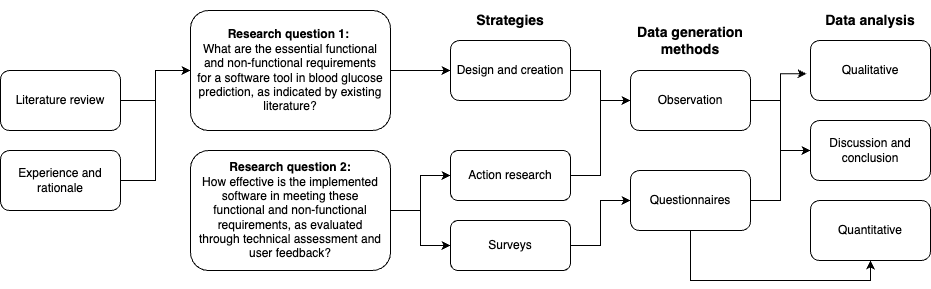}
    \caption{Methodological flowchart for the study.}
    \label{fig:study_process}
\end{figure*}

\section{Methods}

In designing our study, we followed a model for the research process inspired by Oates et al. \cite[p. 34]{OatesResearching2005}. Figure \ref{fig:study_process} illustrates our study process in a flowchart. Our study started with establishing the objectives and design of the GluPredKit platform, drawing from current literature, experience, and logical reasoning, as detailed in \ref{design_and_implementation}. Subsequently, we implemented the software according to these specifications and conducted preliminary tests to ensure its functionality. For validation purposes, we engaged users in testing the platform, enabling them to explore its different features. The method for data collection was a questionnaire incorporating the SUS and open-ended questions. Finally, we analyzed the questionnaire responses to evaluate the platform's effectiveness.

\subsection{Design and Implementation}
\label{design_and_implementation}

The system requirements were designed based on the observed paradigm for BG prediction in \ref{related_work}, with the commonalities and differences between model approaches and model validation, as well as addressing research gaps. 

We categorized the requirements into functional and non-functional aspects. Functional requirements are specific demands for functional primitives, outlining specific features of the app. Non-functional requirements define how well the functions should perform \cite{altexsoft2023functional}.

\subsubsection{Functional Requirements}

The functional requirements, detailed in \textit{Functional Requirements} in supplementary materials, were derived from our literature review. Our analysis revealed a consistent methodology across various studies, encompassing four core stages: data acquisition, preprocessing, model training, and evaluation using metrics and visual tools. To accommodate the variability observed in these studies, GluPredKit was designed with user-configurable settings, such as the choice between $mg/dL$ and $mmol/L$ for BG units (functional requirement \#10).

GluPredKit, as outlined in functional requirement \#11, was designed to foster open-source collaboration and community contributions through a modular architecture. This structure enables a flexible "plug-and-play" approach for the data parsers, preprocessors, model architectures, evaluation metrics, and plots. Additionally, in alignment with requirements \#1 to \#6, we incorporated some components into each module.

\begin{table}[!ht]
    \centering
    \begin{tabular}{p{0.15\textwidth}p{0.7\textwidth}}
        \toprule 
        Modules 
        & 
        Implemented Components 
         \\
        
        \midrule 
        \midrule 
        
        Parsers & 
        \begin{minipage}[t]{0.7\textwidth}
            \begin{itemize}
            \setlength{\itemindent}{-1em} 
                \item Nightscout API
                \item Tidepool API
                \item Apple Health 
                \item Ohio T1DM dataset
            \end{itemize} 
        \end{minipage} \\

        \midrule 
        Models &
        \begin{minipage}[t]{0.7\textwidth}
            \begin{itemize}
            \setlength{\itemindent}{-1em} 
                \item Linear Regression
                \item Elastic Net 
                \item Gradient Boosting Trees
                \item Huber 
                \item Lasso 
                \item Random Forest
                \item Ridge 
                \item Linear Support Vector Regression (SVR)
                \item Radial Basis SVR  
                \item Vanilla Long Short-Term Memory
                \item Temporal Convolutional Network \cite{BaiKolterKoltun2018}
            \end{itemize}
        \end{minipage} \\

        \midrule 
        Metrics & 
        \begin{minipage}[t]{0.7\textwidth}
            \begin{itemize}
            \setlength{\leftmargin}{-1em} 
            \setlength{\itemindent}{-1em} 
                \item Root mean squared error (RMSE)
                \item Mean absolute error (MAE)
                \item Mean absolute relative difference
                \item Mean error
                \item Mean relative error
                \item Glucose-specific RMSE \cite{del_favero_glucose-specific_2012} \cite{jacobs_artificial_2023}
                \item Clarke error grid 
                \item Parkes error grid
            \end{itemize}
        \end{minipage} \\

        \midrule 
        Plots & 
        \begin{minipage}[t]{0.8\textwidth}
            \begin{itemize}
            \setlength{\itemindent}{-1em} 
                \item Scatter plot
                \item Trajectories
                \item Single prediction (with real-time option)
            \end{itemize}
        \end{minipage} \\
        
        \bottomrule 
    \end{tabular}
    \caption{Overview of the implemented components in GluPredKit's modules. This open-source platform invites future contributions and expansions from the community. }
    \label{tab:components}
\end{table}

The complete list of our implemented components in each module in GluPredKit are listed in Table \ref{tab:components}. In deciding which data sources to integrate into the parsers module, we prioritized sources containing several relevant data types so that each source could provide a complete dataset. We identified various sources, including web APIs and data exports that integrate with a broad range of diabetes equipment, as well as the benchmark dataset Ohio T1DM. 

We used the identified prediction models from a comprehensive benchmark study by Xie et al. \cite{xie_benchmarking_2020}. They identified commonly used models based on linear- and nonlinear machine-learning regression techniques and deep-learning algorithms. The implemented components in GluPredKit can provide as a benchmark for new models presented in the future. 

Recently, Jacobs et al. did a consensus study about evaluating BG prediction algorithms, which formed the basis for our decided implemented evaluation metrics \cite{jacobs_artificial_2023}. RMSE, often a primary metric in conjunction with relative error measures like MAE, is a standard evaluation approach in machine learning. However, these metrics may fall short in capturing the clinical significance of BG predictions, as they uniformly penalize errors across the glucose range. To address this limitation, the glucose-specific RMSE has been proposed \cite{del_favero_glucose-specific_2012} \cite{jacobs_artificial_2023}. Moreover, Clarke or Parkes error grid analysis, initially designed for assessing BG measurements \cite{benesch_error_undefined}, has evolved into widely adopted metrics for evaluating the clinical applicability of BG prediction algorithms. These error grid zones are defined based on actual measurements rather than predictions in future time but serve as a valuable supplementary tool for extending the assessment using existing metrics.

Our scatter plot, aligning with prior studies \cite{xie_benchmarking_2020}, compares predicted and measured values. In addition, we introduce two novel plots: The graph in the bottom of Figure \ref{fig:glupredkit} is an example of a plot of predicted BG trajectories generated using GluPredKit. The black dots indicate CGM measurements, while the dashed lines from each measurement point demonstrate the model's predictions over time. Forecasted trajectories over time, in contrast to single-horizon predictions, are crucial for MPC \cite{SpringerMPC2021}. MPC is a common control approach in artificial pancreas systems \cite{moon_current_2021}. The last plot focuses on a single predicted trajectory, illustrating real-time responses to events like meals or insulin injections, offering insights into the model's adaptability and clinical potential.

\begin{figure*}[!ht]
    \centering
    \includegraphics[width=0.8\linewidth]{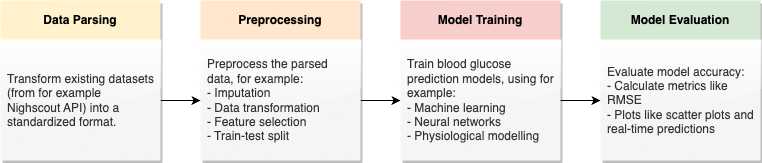}
    \caption{Overview of the GluPredKit pipeline. This diagram delineates the sequential stages involved in the BG prediction process, starting from data parsing from existing datasets, through preprocessing and model training, culminating in model evaluation and real-time predictions.}
    \label{fig:pipeline}
\end{figure*}

\subsubsection{Non-Functional Requirements}

\paragraph{Usability.} The interface of GluPredKit shall be intuitively designed to ensure ease of access to its functionalities. While a fundamental understanding of programming and the terminologies associated with BG prediction is assumed, the system will be accompanied by a command line interface (CLI) and comprehensive documentation. 

\paragraph{Flexibility.} GluPredKit shall exhibit a high degree of flexibility, accommodating the integration of new components by community efforts. Detailed guidelines and protocols for adding such contributions will be systematically documented.

\paragraph{Maintainability.} The codebase of GluPredKit shall be developed with long-term maintainability in mind. It will be structured to allow future developers, even beyond the duration of this study, to manage, modify, and enhance the system efficiently.

\subsection{User Feedback}

\subsubsection{Pilot Testing}

To ensure the robustness and efficacy of our software and documentation, we undertook multiple rounds of pilot testing. In an iterative process, this involved the authors and colleagues until we ascertained a seamless system operation devoid of errors or ambiguities. Participants engaged in a user study (see \textit{User Study} in supplementary materials), which required them to explore the software's features and complete a survey. Feedback garnered during these sessions, especially initial impressions, proved instrumental in implementing enhancements to both the software and its supporting documents.

\subsubsection{User Testing}

The user testing phase assessed the platform's usability, architectural integrity, and overall efficacy. The test was designed to ascertain whether GluPredKit meets the needs of new researchers in the field by balancing user-friendliness with essential functionality. Participants underwent a three-step process: (1) Engage with GluPredKit's functionalities, (2) Complete the SUS questionnaire, and (3) Respond to open-ended questions for additional feedback. The entire questionnaire is provided in \textit{User Study} in supplementary materials. 

\paragraph{Participants.} The study involved four participants, selected based on their interest in BG prediction and proficiency in coding. None had prior exposure to GluPredKit. Although the sample size may appear limited for the quantitative evaluation of SUS scores, the specificity of our inclusion criteria naturally limits the participant pool. Furthermore, the open-ended questions were expected to yield some insights into the software's effectiveness despite the smaller sample size.

\paragraph{Part 1 - Testing GluPredKit Functionalities} Participants received instructions to explore various features of GluPredKit, utilizing a sample dataset in .csv format. The usability test included tasks such as data preprocessing, algorithm training, testing, and comparison, all performed using the platform's command-line interface.

\paragraph{Part 2 - SUS} The SUS, a widely recognized industry standard for usability assessment, consists of ten questions \cite{BrookeSUS1996}. Participants rated each question on a scale of 1 to 5. 

\paragraph{Part 3 - Open-Ended Questions} This section aimed to gather insights into participants' background knowledge and perceptions of the platform's flexibility, particularly regarding the code architecture. Questions also focused on the adequacy of documentation, usability, and the most valuable features encountered. Participants were asked to hypothetically estimate time savings compared to other software, acknowledging the absence of similar existing software for direct comparison. Finally, open-ended questions solicited feedback on potential improvements and the anticipated impact of GluPredKit, informing future development and gauging general interest in the software.


\section{Results}

This section addresses Research Question 2 by assessing how GluPredKit, our developed software, fulfills the established functional and non-functional requirements. Initially, we provide an overview of GluPredKit, followed by an evaluation against our predefined criteria. Lastly, we delve into the insights from the user testing, analyzing responses gathered through questionnaires.

\subsection{GluPredKit - System Description}

GluPredKit consists of an open-source Python repository that facilitates the whole pipeline of collecting new data, preprocessing the data, training a model, and evaluating and comparing prediction models with statistical and graphical representations. The pipeline is illustrated in Figure \ref{fig:pipeline}. A CLI is implemented for users to interact with the system. The code is uploaded to PyPi to be installed with only one command. In this section, we will describe the main components of the system and outline how the system meets the functional and non-functional requirements in the previous section.

Detailed installation guidelines can be accessed from the \href{https://pypi.org/project/glupredkit/0.1.7/}{PyPi documentation} or the project's  \href{https://github.com/miriamkw/GluPredKit/tree/jan_24}{GitHub repository}.

\subsubsection{System Architecture}
GluPredKit's system architecture is a collection of modules specifically tailored to a facet of the BG prediction pipeline, as visualized in Figure \ref{fig:pipeline}. The architecture is divided into distinct modules: parsers, preprocessors, models, metrics, and plots. Each of these has a foundational base model from which specific classes derive. These foundational classes define function requirements and expected input and output formats.

\subsubsection{Command Line Interface}

\begin{figure}
    \centering
    \includegraphics[width=1\linewidth]{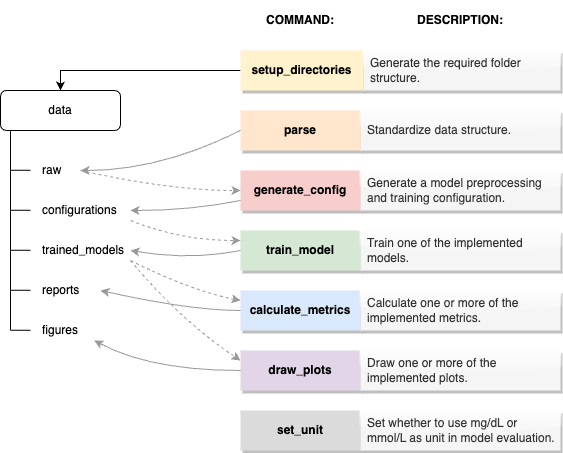}
    \caption{Overview of the CLI commands and how they interact with the file structure in GluPredKit.}
    \label{fig:cli}
\end{figure}

As illustrated in Figure \ref{fig:cli}, GluPredKit provides a CLI that includes a suite of commands for managing the end-to-end data processing and model evaluation workflow. The figure also depicts the file structure in GluPredKit and the file flow for organizing the processed data, model outputs, and evaluation results.

The GluPredKit's functionality is initiated by the \texttt{setup\_directories} command, which constructs a data folder with necessary subfolders. Then, the parse command can be called, producing a raw dataset and storing it within the \texttt{data/raw} directory, as shown in the figure. This dataset forms the foundation for subsequent commands. For instance, the \texttt{generate\_config} command relies on the raw dataset to create a preprocessing and training configuration. Each step in the workflow corresponds with specific directories and files within the system's file structure, ensuring an organized and efficient data management environment.

\subsubsection{Implementation Details}

GluPredKit has primarily been developed using Python. We have utilized GitHub for version control and for hosting the code in an open-source manner. GluPredKit is available on PyPi, facilitating the use of the CLI directly or integrating GluPredKit as a sub-component within another system. The CLI is implemented using the click library.

In our endeavor to implement BG prediction models, we have used multiple libraries. For instance:

\begin{itemize}
                \item Vanilla Long Short-Term Memory
    \item Keras, built atop TensorFlow, facilitates training our Deep Neural Networks (DNN), specifically Long Short-Term Memory and Temporal Convolutional Network models.
    \item PyTorch, also a DNN library.
    \item XGBoost is employed for gradient-boosting trees.
    \item For the remaining machine learning models, we have utilized scikit-learn.
\end{itemize}

 \subsubsection{Reproduction of Benchmark Results}

To validate that our data processing and model training accurately were reproducing the steps described in \cite{xie_benchmarking_2020}, we acquired the same dataset that they used, namely the Ohio T1DM dataset and compared our results. \textit{Reproduction of Results} in supplementary materials provides our reproduced tables, figures, and discussion of the reproduced results.

\subsection{Fulfillment of System Requirements}

The development of GluPredKit has been aligned with the pre-established functional and non-functional requirements in the following ways:

\subsubsection*{Functional Requirements}
\begin{itemize}
    \item Requirement \#1 to \#10 are met through various CLI commands: \begin{itemize}
    \item \texttt{parse}: Dataset retrieval and standardization.
    \item \texttt{generate\_config}: User-specific requirements to the predictive modeling.
    \item \texttt{train\_model}: Model training.
    \item \texttt{calculate\_metrics} and \texttt{draw\_plots}: Model evaluation.
    \item \texttt{set\_unit}: Unit configuration.
    \end{itemize}
    \item Requirement \#11: Achieved through GluPredKit's modular design, which encourages community contributions by offering a standardized interface for all modules.
\end{itemize}

\subsubsection*{Non-Functional Requirements}

\begin{itemize}
    \item \textbf{Usability: } Fulfilled by the system's user interface, detailed documentation, and a demonstrative video.
    \item \textbf{Flexibility: } A modular design underpins the system architecture, enhancing adaptability for integrating new components. Clear base classes and extensive documentation fortify this modularity.
    \item \textbf{Maintainability: } Our commitment to maintainability is underpinned by steadfast adherence to consistent language, alignment with established paradigms, and thorough documentation. 
\end{itemize}

\subsection{User Test Results}

\subsubsection{SUS Results}

\begin{figure}
    \centering
    \includegraphics[width=1\linewidth]{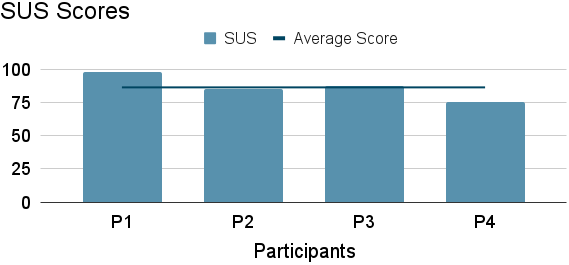}
    \caption{Bar chart of individual SUS scores alongside a line indicating the average SUS score.}
    \label{fig:sus_results}
\end{figure}

Figure \ref{fig:sus_results} visualizes the SUS scores for each participant, revealing an impressive average of 86 (SD = 9). Interpreting these scores, as per \cite{Sauro2018InterpretSUS}, our system garners an 'A' grade in usability on average, though one participant falls into the 'B' range. Descriptively, these scores align with 'Best Imaginable' to 'Good' user experiences, indicating high usability satisfaction among the participants.

\subsubsection{Open-Ended Questionnaire Insights}

\paragraph{Feedback on Software and Documentation}
Participants generally praised the documentation's thoroughness, although some noted challenges in data parsing and error feedback. Suggestions for improvement included more intuitive data parsing guides and more apparent error indications.

\paragraph{Recommendations for System Enhancement}
Respondents recommended further documentation enhancements, flexibility in DNN configuration, expanding data source compatibility, and adding automated hyperparameter tuning. Some suggested developing a user interface to broaden accessibility beyond technically proficient users.

\paragraph{Impact of GluPredKit}
Participants anticipated significant time savings using GluPredKit, with estimations like “10x longer without it” for comparable tasks. They emphasized GluPredKit's potential in advancing diabetes management through its standardized approach and the possibility of community-driven improvements and educational benefits. Highlights include:

\begin{quote}
``Its standardized approach and ease of integrating and comparing new models can significantly enhance prediction accuracy and efficiency in diabetes management.''
\end{quote}

and

\begin{quote}
``The most impressive part is that it encourages code contributions and the establishment of a community around the software for continuous improvement. [...] I also think it has an educational value. The software, through its documentation and examples, can serve as an educational resource for individuals entering the field of BG prediction [...].''
\end{quote}


\section{Discussion}

In this study, we aimed to create GluPredKit, a specialized software for BG prediction and evaluation. This platform was specifically designed to integrate a wide range of predictive methodologies and algorithms. Our development strategy was informed by insights from a benchmark study \cite{xie_benchmarking_2020}, and we concentrated on incorporating several advanced algorithms. The effectiveness of these algorithms was validated by replicating results from the benchmark study, as detailed in \textit{Reproduction of Results} in supplementary materials.

Addressing our first research question, we identified essential functional and non-functional requirements for a BG prediction tool based on a literature review and our practical experiences. These requirements, listed in \textit{Functional Requirements} in supplementary materials, cover the entire prediction model pipeline, including data acquisition, preprocessing, training, evaluation, flexible user settings, and system adaptability.

For our second research question, we evaluated GluPredKit through technical assessments and user feedback. The technical assessment confirmed that GluPredKit successfully meets all our predefined system requirements. To further assess its efficacy and utility, we conducted a user study involving four participants who engaged with all features of GluPredKit and then provided feedback through a questionnaire. The software's usability was quantitatively affirmed by an average SUS score of 86 (SD = 9), indicating high user satisfaction. The study's qualitative feedback from open-ended questions underscored GluPredKit's effectiveness in addressing standardization challenges and its educational benefits, enriched by comprehensive documentation. Participants also offered constructive suggestions for further enhancements, such as developing a graphical user interface, refining data parsing documentation, and expanding the modular framework to include more components.

The field of BG prediction, as noted by Jacobs et al. \cite{jacobs_artificial_2023}, faces a challenge due to the need for uniformity in methodologies, evaluation metrics, and data usage. This inconsistency hampers research progression, as new models are often compared against varying standards. GluPredKit addresses these issues by promoting open-source code, standardizing critical stages like data parsing and preprocessing, and establishing consistent evaluation metrics. Moreover, GluPredKit can be used to train existing algorithms, and to evaluate them in real-world settings, a research gap identified by Diouri et al. \cite{diouri_hypoglycaemia_2021}.

If widely adopted, GluPredKit has the potential to streamline the benchmarking of new algorithms against established methods. Its modular architecture, designed around a "plug-and-play" framework, ensures compatibility with various datasets, models and evaluation metrics. GluPredKit is intended for collaborative efforts, where researchers can employ GluPredKit for iterative refinement of their predictive models. This collaborative effort involves conducting evaluations across diverse datasets, and making incremental improvements. If a researcher notices that their model consistently outperforms others compared to existing benchmarks, they can submit a pull request, contributing their enhanced model to the academic community. This iterative and collaborative approach could elevate the quality of predictive models, where GluPredKit serves as a hub for innovation and benchmarking within the academic landscape.

However, our study has limitations, notably the small number of participants in the user study, which may only partially capture the broader range of potential users. Furthermore, replicating existing studies poses inherent challenges, although our results closely align with established findings. A key design goal for GluPredKit was to balance flexibility with simplicity, avoiding unnecessary complexity.

\section{Conclusion}

This study aimed to develop and assess GluPredKit, a software platform to standardize the process of BG prediction and facilitate comparative analysis. We outlined a pipeline for BG prediction, encompassing data acquisition, preprocessing, model training, and evaluation. GluPredKit, with its modular architecture and standardized input and output formats, encourages community contributions. It demonstrates satisfactory usability as an open-source tool supplemented with a sample dataset, CLI, and video tutorial. The user study, though limited in participant number, supports the platform's efficacy. Future initiatives will focus on integrating additional algorithms into GluPredKit, encompassing physiological models, machine learning techniques, and DNNs. Additionally, more user feedback shall be collected to guide further development. Lastly, future work should address the improvements proposed in this user study and validate algorithms in real-world scenarios.

\bigskip

\bibliographystyle{plain}
\bibliography{references}

\end{document}